\documentclass[twocolumn,showpacs,preprintnumbers,amsmath,amssymb,prb,floatfix,superscriptaddress]{revtex4}

\usepackage{graphicx,tabularx}
\usepackage{amssymb}
\usepackage{dcolumn}
\usepackage{color}
\usepackage[mathcal]{euscript}
\usepackage{color}
\usepackage{multirow}
\definecolor{gray0}{gray}{0.0}
\definecolor{gray64}{gray}{0.25}
\definecolor{gray128}{gray}{0.5}
\definecolor{gray192}{gray}{0.75}
\definecolor{gray255}{gray}{1.0}

\def\wH{$w^H_{A}(\mathbf{r})$}

\begin{document}

\title{Extending Hirshfeld-I to bulk and periodic materials.}
\author{D. E. P. Vanpoucke}
\affiliation{SCRiPTS group, Dept. Inorganic and Physical Chemistry, Ghent University, Krijgslaan $281$ - S$3$, $9000$ Gent, Belgium}
\affiliation{Ghent Quantum Chemistry Group, Dept. Inorganic and Physical Chemistry, Ghent University, Krijgslaan $281$ - S$3$, $9000$ Gent, Belgium}
\author{P. Bultinck}
\affiliation{Ghent Quantum Chemistry Group, Dept. Inorganic and Physical Chemistry, Ghent University, Krijgslaan $281$ - S$3$, $9000$ Gent, Belgium}
\author{I. Van Driessche}
\affiliation{SCRiPTS group, Dept. Inorganic and Physical Chemistry, Ghent University, Krijgslaan $281$ - S$3$, $9000$ Gent, Belgium}

\date{\today}
\begin{abstract}
In this work, a method is described to extend the iterative Hirshfeld-I method, generally used for molecules, to periodic systems. The implementation makes use of precalculated pseudo-potential based charge density distributions, and it is shown that high quality results are obtained for both molecules and solids, such as ceria, diamond, and graphite. The use of such grids makes the implementation independent of the solid state or quantum chemical code used for studying the system. The extension described here allows for easy calculation of atomic charges and charge transfer in periodic and bulk systems.
\end{abstract}

\pacs{  } 
\maketitle
\section{Introduction}
\indent One of the most successful concepts in chemistry is that of ``atoms in molecules'' (AIM). It states that the properties of a molecule can be seen as simple sums of the properties of its constituent atoms. An impressive amount of insights has been obtained from such a viewpoint, although a precise definition of an AIM remains elusive.\cite{AyersPW:2000JCP,BaderRFWMattaCF:2004JPhysChemA,BaderRFWMattaCF:2006JPhysChemA}
All AIM methods have the common purpose of trying to improve our understanding of chemical concepts such as molecular similarity and transferability between molecules.\cite{ParrRGAyersPWNalewajskiRF:2005JPCA}
Since the concept of AIM is basically about how one should divide the electrons, more specifically the charge density distribution (CDD) $\rho(\mathbf{r})$ in the molecule between the different ``atoms'', this leads to two obvious categories of approaches in which most of the methods used for defining AIM can be divided. The first category of approaches is based on the wave-function/states of the system, and most of the work is performed in the Hilbert space of the basis functions used. One of the most famous examples here is (probably) the Mulliken approach.\cite{MullikenRS:1935JChemPhys,MullikenRS:1955aJChemPhys}
The second category of approaches is based on the division of the CDD as it exists in real-space space. In these real-space approaches the molecule is split into atomic basins, that can overlap such as in the Hirshfeld\cite{HirshfeldFL:1977TCA} and derived methods,\cite{BultinckP:2007FaradDisc_HirRef,BultinckP:2007JCP_HirRef,LillestolenTCWheatleyRJ:2008ChemComm,LillestolenTCWheatleyRJ:2009JChemPhys,ManzSholl:JChemTheoryComp2010} or that are non-overlapping such as in Bader's approach.\cite{BaderRFW:1991ChemRev}\\
\indent The concept of AIM is strongly linked to the concept of transferability. Because both are central in chemistry, and chemists mainly focus on molecules, they are mostly used for molecules.\cite{FranciscoEPendasAM:2011CTC,FranciscoEPendasAM:2011TCA} There is, however, no reason why these concepts should not be applicable for periodic systems such as bulk materials. Even more, if these concepts are truly valid, they should hold equally well for solids as for molecules, and should provide additional insight in the chemical properties of defects, such as dopants, interfaces and adsorption of molecules on surfaces.\\
\indent In this work, we have implemented an extension of the iterative Hirshfeld-I approach\cite{BultinckP:2007JCP_HirRef} to periodic systems, as a module in our HIVE-code.\cite{HIVE_REFERENCE} The implementation makes use of grid stored CDDs, which can easily be generated by standard solid state and quantum chemical codes.\\
\indent In Sec.~\ref{sc:Hir_Methods} a short review of the parameters used in the solid state and quantum chemical codes is given. Afterwards the basic theory behind the Hirshfeld-I method is presented and extended to periodic systems. In addition, the spatial integration grid, pseudo-potentials and stored CDDs are discussed in view of the Hirshfeld-I method for periodic systems. In Sec.~\ref{sc:Hir_Res_and_Dis} the influence of the different grids on the accuracy is discussed. We also identify a delocalization problem in the radial CDDs which originates from the plane wave approach and periodic boundary conditions (PBC) used in the solid state code, and a simple and effective solution for this issue is presented. As a last point, the atomic charges in some simple periodic systems are calculated, showing that the algorithm works correctly. Finally, in Sec.~\ref{sc:Hir_conclusions} some conclusions are given.\\
\section{Methods}\label{sc:Hir_Methods}
\subsection{Atomic and molecular calculations}\label{ssc:Hir_Methods_solstatecodes}
\indent Hirshfeld-I calculations require CDDs as input. These can be obtained from electronic structure calculations using standard solid state or quantum chemical codes. In this work we have chosen to perform these calculations within the DFT framework using the projector augmented wave (PAW) approach for the core-valence interaction and the local density approximation (LDA) for the exchange-correlation functional as implemented in the \textsc{VASP} code.\cite{Blochl:prb94,Kresse:prb99} The kinetic energy cut-off is set at $500$ eV and the $k$-point set is reduced to the  $\Gamma$-point for molecular and atomic calculations. For the bulk materials sufficiently converged $k$-point sets were used. To optimize the geometry of the molecules and periodic materials a conjugate gradient algorithm is applied. For molecular calculations only the atom positions are optimized, for bulk materials the cell parameters are optimized simultaneously. All molecules are placed in periodic cells of $20.0\times 21.0\times 20.5$\AA$^3$, which provide a sufficiently large vacuum region between periodic copies of the molecules, to prevent interaction.\\
\indent Hirshfeld-I data computed using the approach detailed in the present study, are compared to those obtained using more common molecular calculations of AIM properties. For the set of $168$ neutral molecules previously studied by Bultinck \textit{et al.},\cite{BultinckP:2007JCP_HirRef,VanDamme_Bultinck:JCTC2009} geometry optimization and Hirshfeld-I charge calculations are performed at the Local Spin Density Approximation\cite{HohenbergKohn:PR1964,KohnSham:PR1965} level with the Slater exchange functional\cite{SlaterJC:1974} and the VWN$5$ correlation functional\cite{VoskoWilkNusai:CanJP1980} as implemented in Gaussian-03.\cite{Gaussian03} Numerical integrations are carried out using Becke's integration grid with $170$ angular points in the Lebedev-Laikov grid.\cite{BeckeAD:1987JCP,LebedevVI_grid:1999DokladyMath} The Hirshfeld-I charges are considered converged if the largest change in charge of any atom is below $0.0005$e.

\subsection{Hirshfeld methods}
\indent The basic idea behind the Hirshfeld method,\cite{HirshfeldFL:1977TCA} also known as the stockholder method, is that the AIM share the charge density in each point of space. This means that the AIM CDD becomes a weighted partition of the molecular CDD. Formally this can be written as:
\begin{eqnarray}
\rho_{mol}(\mathbf{r})&=&\sum_{A}{\rho_A^{AIM}(\mathbf{r})}\nonumber\\
\rho_A^{AIM}(\mathbf{r})&=&w_{A}^{H}(\mathbf{r})\rho_{mol}(\mathbf{r}) \qquad \forall \mathbf{r}\\
\mathrm{with\ }\sum_{A}{w_{A}^{H}(\mathbf{r})} &\equiv& 1 \nonumber,
\end{eqnarray}
with $\rho_{mol}(\mathbf{r})$ and $\rho_{A}^{AIM}(\mathbf{r})$ the CDDs for the molecule and the AIM. All sums are taken over the entire set of AIM, and \wH\ is the Hirshfeld weight function for atom $A$. From these equations the Hirshfeld weight can be written as:
\begin{equation}
w_{A}^{H}(\mathbf{r})=\frac{\rho^{AIM}_{A}(\mathbf{r})}{\sum_{B}{\rho_{B}^{AIM}(\mathbf{r})}}.\label{eq:Hir_w_mol}
\end{equation}
Since the $\rho^{AIM}(\mathbf{r})$ are the CDDs sought, they cannot be used as input. The Hirshfeld method circumvents this problem by using spherically averaged reference state atomic CDDs $\rho_{X}^{\circ}(\mathbf{r})$. In the original paper by Hirshfeld the neutral atomic ground state is used as reference state.\cite{HirshfeldFL:1977TCA} When summing these isolated atomic CDDs over all AIM, one gets the so-called `\emph{promolecular}' CDD instead of the actual molecular CDD:
\begin{equation}\label{eq:promoldef}
\rho_{promol}(\mathbf{r})=\sum_{B}{\rho^{\circ}_{B}(\mathbf{r})}.
\end{equation}
It is then assumed that the difference between this promolecular CDD and the actual molecular CDD has only little influence on the Hirshfeld weight \wH. As a result one can write the CDD of an AIM $A$ as:
\begin{equation}\label{eq:Hir_chg_AIM}
\rho^{AIM}_{A}(\mathbf{r})=\frac{\rho^{\circ}_{A}(n_{A}^{\circ},\mathbf{r})}{\rho_{promol}(\mathbf{r})}\rho_{mol}(\mathbf{r})=w^{H}_{A}(\mathbf{r})\rho_{mol}(\mathbf{r}),
\end{equation}
where the population of the atom $A$ is given by $n_{A}^{\circ}$. In the original Hirshfeld approach, neutral atoms were used as reference. This, however, has been identified by several authors to be a major weakness of the method as changing the choice of the promolecular atom charges can have a highly significant effect on the resulting AIM\cite{BultinckP:2007JCP_HirRef,BaderRFWMattaCF:2004JPhysChemA,BaderRFWMattaCF:2006JPhysChemA,DavidsonERChakravortyS:1992TheorChimActa,FranciscoEPendasAM:2007JPhysChemA}
From eq.\eqref{eq:Hir_chg_AIM} it is easy to understand that the resulting $\rho^{AIM}_{A}(\mathbf{r})$ will tend to be as similar to $\rho^{\circ}_{A}(n_{A}^{\circ},\mathbf{r})$ as possible\cite{AyersPW:2000JCP,AyersPW:2002JCP,BultinckP:2007JCP_HirRef}, explaining why the Hirshfeld populations strongly depend on the choice of reference atomic CDDs. Fortunately, this problem can be resolved by using the iterative Hirshfeld-I scheme.\cite{BultinckP:2007FaradDisc_HirRef,BultinckP:2007JCP_HirRef} For each iteration $i$, the obtained $\rho^{AIM}_{A}(\mathbf{r})$ are used to calculate the population $n^{i}_{A}$ of each atom $A$. The (spherically symmetric) CDD $\rho^{i}_{A}(n_{A}^{i},\mathbf{r})$ of a free atom $A$ with population $n^{i}_{A}$ is then used as atomic CDD in \wH. For each iteration the new promolecular density $\rho^{i}_{promol}(\mathbf{r})$ is  obtained by summing the density distributions $\rho^{i}_{A}(n_{A}^{i},\mathbf{r})$ for all atoms of the molecule. This setup is independent of the initial choice of atomic CDDs and the convergence of the iterative scheme is determined by the convergence of the populations of the AIM.\cite{BultinckPAyersPWFiasS:2007ChemPhysLett,GhillemijnDBultinckP:2011JCompChem} Note that the first step in this scheme usually corresponds to the standard Hirshfeld method with $n_{X}^{\circ}=Z_{X}$. At this point it is important to note that the CDDs $\rho^{AIM}_{A}(\mathbf{r})$ and $\rho_{A}^{i}(n_{A}^{i},\mathbf{r})$ will generally be different, despite having the same population. $\rho_{A}^{i}(n_{A}^{i},\mathbf{r})$ is constructed as a spherically symmetric CDD, whereas $\rho_{A}^{AIM}(\mathbf{r})$ is a weighted part of the molecular CDD. The resulting CDD will generally be not spherically symmetric but show protrusions along the directions bonds are formed.\\
\indent The extension of the Hirshfeld and Hirshfeld-I methods from molecules to bulk and other periodic materials is quite trivial from the formal perspective. The main problem lies in the fact that a bulk system is considered to consist of an infinite number of ``\emph{atoms in the system}''(AIS). Calculating the atomic charge densities for all AIS can, for a periodic system, be reduced to only the atoms in a single unit cell since all periodic copies should yield the same results.\\
\indent In addition to this, also the summation limits in  Eqs.~\eqref{eq:Hir_w_mol},~\eqref{eq:promoldef} and \eqref{eq:Hir_chg_AIM} change. Where for molecules the sum over $B$ is a finite sum over all AIM, it becomes an infinite sum over all AIS. Because atomic CDDs drop exponentially, the density contribution to the ``prosystem'' CDD $\rho_{prosys}(\mathbf{r})=\sum^{AIS}_{B}{\rho_B^{\circ}(n_{B}^{\circ},\mathbf{r})}$ of atoms at larger distances becomes negligible. This allows us to truncate the infinite sum to include only the atoms within a certain `\emph{sphere of influence}' (SoI), \textit{i.e.} all atoms of which the contribution to the prosystem CDD is not negligible. Within the iterative Hirshfeld-I scheme we then get:
\begin{equation}\label{eq:Hir_w_iterativeHirshfeldDef}
w^{H,i}_{A}(\mathbf{r}) = \frac{\rho^{i-1}_{A}(n_{A}^{i-1},\mathbf{r})}{\sum_{B}^{SoI}\rho_{B}^{i-1}(n_{B}^{i-1},\mathbf{r})} \quad \forall A\ \epsilon\ \mathrm{unit\ cell}
\end{equation}
where
\begin{equation}
\sum_{A}^{SoI}w^{H,i}_{A}(\mathbf{r})\leq \sum_{A}^{AIS}w^{H,i}_{A}(\mathbf{r})\equiv 1 \qquad \forall \mathbf{r},
\end{equation}
with $i$ indicating the iteration step, and $\rho^{i}_{A}(n_{A}^{i},\mathbf{r})$ the atomic CDD for an atom $A$ with a population $n_{A}^{i}$ given by
\begin{equation}\label{eq:Hir_pop_int}
n_{A}^{i} = \int{w^{H,i}_A(\mathbf{r})\rho_{system}(\mathbf{r})\mathrm{d}\mathbf{r}},
\end{equation}
where $\rho_{system}(\mathbf{r})$ is the CDD of the periodic system.

\subsection{Spatial integration of the population}\label{ssc:spatInteg}
\begin{figure}
  \includegraphics[width=8.0cm,keepaspectratio=true]{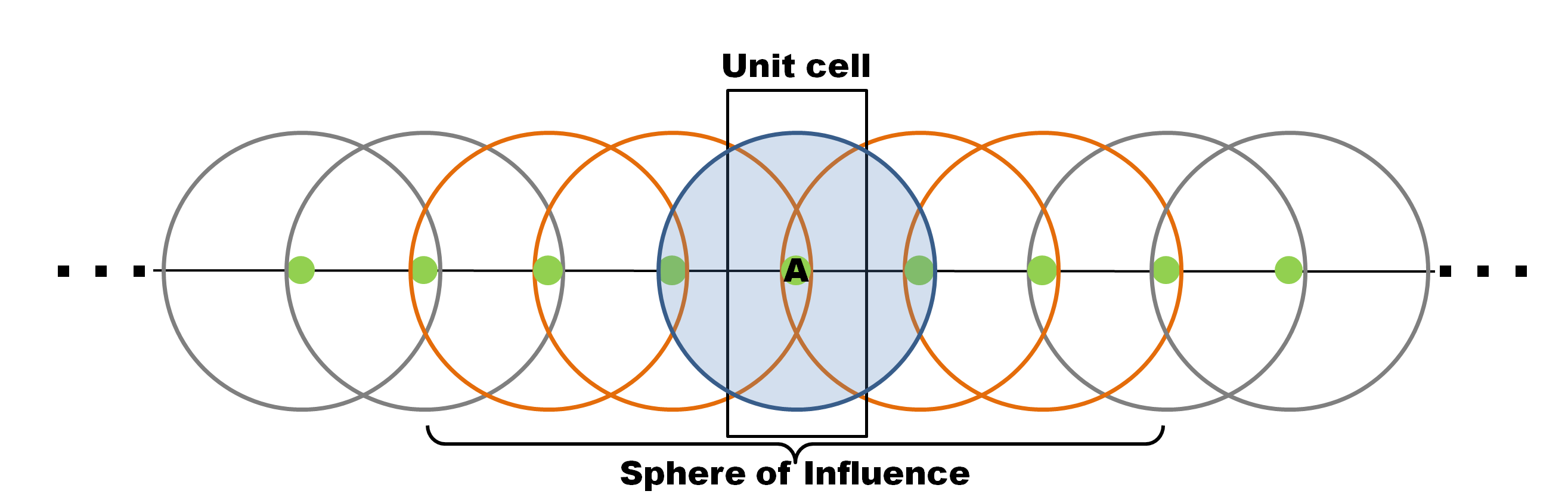}\\
  \caption{(color online) Schematic representation of the integration scheme used for a linear periodic system of atoms (green discs). The black rectangle indicates the unit cell, and the circles indicate the boundaries of the spherical integration regions. The spherical integration regions for the atoms in the sphere of influence (SoI) of atom A are shown in orange, those outside are shown in grey. All integrations can be limited to the blue region (or sections thereof) around atom A (see text). }\label{fig:Hir_SphereOfInfluence}
\end{figure}

\indent In chemistry, due to the exponential decay of the charge density of atoms and molecules, and due to the sharp cusps present in CDDs at the atomic nuclei, atom centered grids are widely and successfully used. This makes them ideally suited for integrations such as given in Eq. \eqref{eq:Hir_pop_int}. The multicenter integration scheme proposed by Becke splits up the full space integration into a set of overlapping atom centered spherical integrations.\cite{BeckeAD:1987JCP} To solve the problem of double counting in the overlapping regions a weight $h_{A}(\mathbf{r})$ is given to each point in space for every atom A in the system, such that
\begin{equation}
\sum_{A}^{AIS}{h_{A}(\mathbf{r})}\equiv 1\qquad \forall\mathbf{r}.
\end{equation}
\indent This weight function indicates how much a point `belongs' to a certain atom $A$. The weight function can be binary, when the space is split up in Voronoi or Wigner Seitz cells,\cite{VoronoiGeorges:JRAM1907,WignerSeitzCell:PhysRev1933} or smoothly varying, as is the case in the Becke scheme.\cite{BeckeAD:1987JCP} As a result, an integrand $F(\mathbf{r})$ can be decomposed as $F(\mathbf{r})=\sum_{A}^{AIS}{h_{A}(\mathbf{r})F_{A}(\mathbf{r})}$ and the full integration becomes
\begin{equation}\label{eq:Becke_IntSplit}
I=\int{F(\mathbf{r})\mathrm{d}\mathbf{r}}=\sum_{A}^{AIS}{\int{h_{A}(\mathbf{r})F_{A}(\mathbf{r})\mathrm{d}\mathbf{r}}},
\end{equation}
where the sum over all AIS is again an infinite sum. However, in numerical implementations for periodic systems, the exponential decay of the atomic CDD allows us to truncate both the infinite sum and integration region of Eq.\eqref{eq:Becke_IntSplit}, without significant loss of numerical accuracy. The sum can be reduced to contain only the atoms included in the SoI of atom $A$ (orange circles in Fig.~\ref{fig:Hir_SphereOfInfluence}), because only these atoms contribute significantly to the density in the integration region around atom $A$. In addition, the integration region for all atoms in the SoI can be reduced even further, without loss of accuracy, to only the region that overlaps with the spherical integration region of atom $A$ (blue shaded disc in Fig.~\ref{fig:Hir_SphereOfInfluence}).

\subsection{Grid stored charge densities and frozen core pseudo-potentials.}
\indent In periodic systems, the use of PBC allows one to reduce the system size dramatically. For bulk materials this even allows simple systems, such as face-centered cubic Cu or Ni, to be represented using single atom unit cells. A useful side effect of such reduced cells is that it is easily possible and relatively cheap to store the CDD $\rho_{system}(\mathbf{r})$ on a three dimensional grid covering the unit cell, and thus fully describing the entire infinite system. The use of such precalculated charge density grids speeds up the Hirshfeld method significantly, since there is no more need to calculate the charge density at any given grid point starting from the wave function of the system. This also makes the implementation independent of the code used to generate the system. The drawback, however, is a slightly reduced accuracy. Since the charge density grid has a finite resolution, interpolation between the stored grid points is needed. This effect is discussed in the following section.\\
\indent In chemistry, it is usually sufficient to consider only the valence electrons to describe the interactions between atoms.
The core electrons are often considered inert as a result, and are kept frozen during calculations, greatly reducing the computational cost for heavy atoms. This has two small but interesting side effects on the calculated radial charge distributions. Firstly, the integrated charge equals the number of valence electrons only, since the core electrons are not explicitly treated. Secondly, the resulting radial profile is not necessarily monotonically decreasing, showing a minimum or even negative values at the core of the atom (\textit{e.g.} Fig.~\ref{fig:radialdens}). The origin of this behavior lies in the practical implementation of the PAW pseudo-potentials.\cite{Blochl:prb94,Kresse:prb99} By using the same pseudo-potentials for generating the atomic charge distributions and the charge distribution of the system under study, however, no errors are introduced since both systems contain the same frozen core contribution.
\section{Results and Discussion}\label{sc:Hir_Res_and_Dis}
\indent To this date we are aware of few implementations for periodic systems of the Hirshfeld and \mbox{Hirshfeld-I} approach, but their number is steadily rising. Pend\'{a}s \textit{et al.}\cite{PendasAM:2002JChemPhys} investigated Hirshfeld surfaces as approximations for interatomic surfaces for LiF and CS$_2$ crystals. The Cut3D plugin of the \textsc{ABINIT} code can be used to calculate Hirshfeld charges,\cite{ABINIT_REFERENCE:ComputerPhysCommun2009,SmithMD:CrystGrowthDes2011,GlorEC:JSolStateChem2011,SchierJ:2011ApplMaterInter} and recently Leenaerts \textit{et al.} implemented a ``subsystem'' based Hirshfeld-I method to study graphane, graphene fluoride and paramagnetic adsorbates on graphene.\cite{LeenaertsOPeetersF:2010PhysRevB,LeenaertsO:AplPhysLett2008,LeenaertsO:PRB2009}
In a recent publication Watanabe \textit{et al.}\cite{WatanabeManzSholl:JPhysChemC2011} presented Hirshfeld results for metal--organic frameworks using the DDEC-code of Manz and Sholl.\cite{ManzSholl:JChemTheoryComp2010} The same code was probably also used in the investigation of charge injection in graphene layers by Rogers and Liu.\cite{RogersGWLiuJZ:JACS} These authors present Hirshfeld-I charges, though they do not mention explicitly how these were obtained.\\
\indent It is clear that there is a growing interest in codes that can provide atomic populations, however, only few true bulk systems have been investigated using a purely atom based Hirshfeld-I method. For this reason most numerical tests in this work are performed on molecules, though we will investigate the behavior of periodic systems at the end of this section. As a first test system we have chosen the CO molecule. Its small size makes it easily suitable for quick test calculations, and its heteronuclear structure should result in a non-zero charge transfer, at least at equilibrium distance.\\
\indent Before proceeding, the different grids used in our current setup of the Hirshfeld-I scheme for periodic systems are introduced. In this setup there are two `types' of CDDs: that of the system and that of the free atoms/ions, which are indicated in the following as the subscripts `$sys$' and `$atom$', respectively. In this, `system' refers to the object of which we want to obtain the atomic charges, and can thus refer to bulk materials, wires, molecules or even single ions. The free atoms/ions on the other hand refer to the single atoms which are used for the generation of the reference radial CDD $\rho_{X}^{i}$ of eqn.\eqref{eq:Hir_w_iterativeHirshfeldDef}. For both types there are two kinds of 3D grids involved:
\begin{enumerate}
  \item Linear grids: Instead of using the analytical expression for the underlying wave function, we use the CDDs stored by \textsc{VASP} on a finite numerical grid. These grids span a single unit cell and use uniformly spaced grid points in direct coordinates.\cite{fn:CHGCAR} (\textit{cf.}~Sec.\ref{ssc:chargedensitygrid}) In the remainder, the following notation is used:
      \begin{itemize}
        \item V$_{atom}$ : The linear grid for the reference atom density distributions as obtained from the atomic calculations.
        \item V$_{sys}$ : The linear grid for the (poly)atomic system under study
      \end{itemize}
  \item Spherical grids: These are atom centered grids which are not limited to a single unit cell. The spherical grids decompose into a radial and a shell grid. In our current setup, a logarithmic grid is used as radial grid, such that closer to the core the grid is sufficiently dense to describe this region accurately. The number of radial points was chosen to equal the numbers suggested by Becke.\cite{BeckeAD:1987JCP} At each point in this radial grid grid, a shell is located on which grid points are distributed according to a Lebedev-Laikov grid.\cite{LebedevVI_grid:1999DokladyMath} (\textit{cf.}~Sec.~\ref{ssc:sphereIntgrid}) The total number of grid points $S$ equals the sum over all atoms of the number of radial points ($R_{A}$) used for that atom times the number of points on each shell ($\sigma$): $S=\sum_{A}{R_{A}\cdot\sigma}$. In the remainder only $\sigma$ is varied to study the stability of the integrations. Two three-dimensional spherical grids are distinguished:
      \begin{itemize}
        \item S$_{atom}$ : Total spherical grid used to generate the reference spherically averaged radial density distribution for the atoms.
        \item S$_{sys}$ : The multi-center spherical grid for the system under study.
      \end{itemize}
\end{enumerate}
\indent Note that the results of the Hirshfeld-I AIM analysis depend directly on $S_{sys}$ but also indirectly on $S_{atom}$ as this determines the quality of the isolated atomic CCDs.

\begin{figure}
  \includegraphics[width=8cm,keepaspectratio=true]{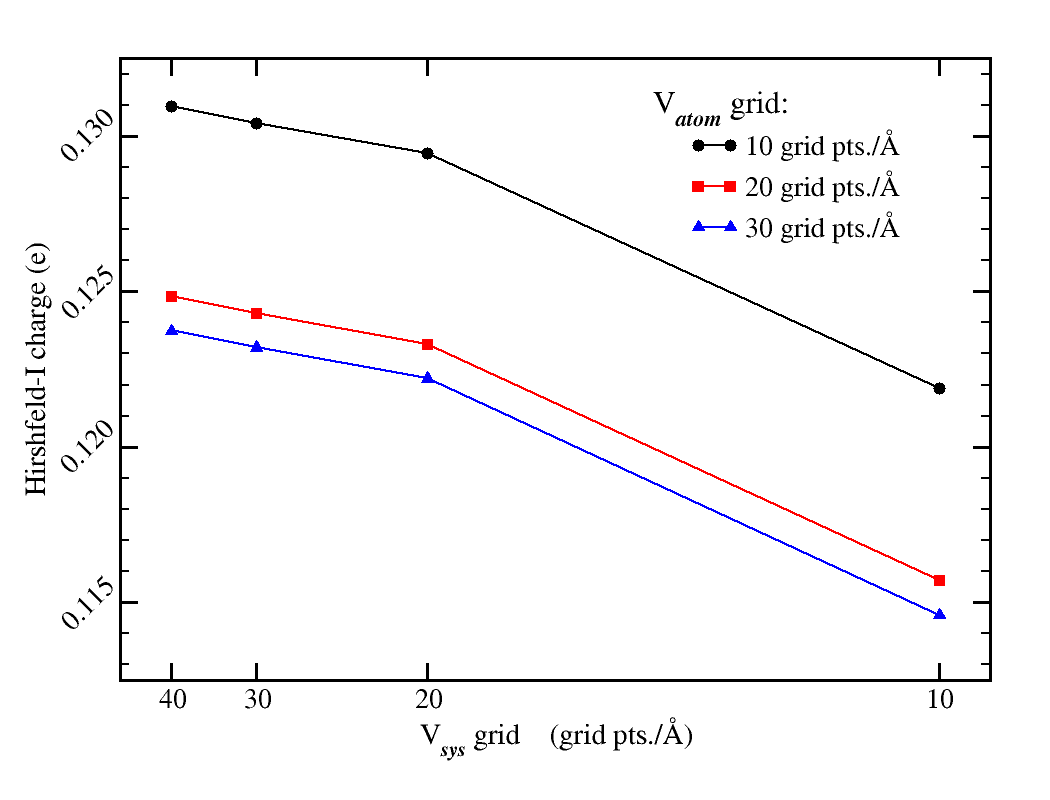}\\
  \caption{The Hirshfeld-I charge for the C atom in a CO molecule as function of the V$_{sys}$ grid resolution. The different curves show the results for the use of different resolutions in the V$_{atom}$ grids, used for generating the atomic radial densities. In all molecular calculations we used spherical integration grids of $1202$ grid points per shell.}\label{fig:HIR_GridConv}
\end{figure}
\begin{table}[tb!]
\caption{Differences in the Hirshfeld-I populations shown in Fig.~\ref{fig:HIR_GridConv}. The presented change in population is the difference in population going from V$_{sys}$/V$_{atom}$ grids with $I$ grid points per \AA, to grids with $J$ grid points per \AA. The respective grid changes are indicated as $\Delta^{I}_{J}$.\\
Top table: Differences along the curves shown in Fig.~\ref{fig:HIR_GridConv}.\\
Bottom table: Differences between the curves shown in Fig.~\ref{fig:HIR_GridConv}. \label{table:gridConv}}
\begin{ruledtabular}
\begin{tabular}{c|ccc}
\multicolumn{1}{c|}{V$_{sys}$} & \multicolumn{3}{c}{V$_{atom}$ (grid points/\AA)}\\
\multicolumn{1}{c|}{ }& $10$ & $20$ & $30$ \\ \hline\\[-2mm]
$\Delta^{10}_{20}$ & 0.00738 & 0.00743 & 0.00744 \\[1.5mm]
$\Delta^{20}_{30}$ & 0.00126 & 0.00128 & 0.00128 \\[1.5mm]
$\Delta^{30}_{40}$ & 0.00049 & 0.00048 & 0.00049 \\[1mm]
\hline
\hline
\multicolumn{1}{c|}{V$_{sys}$} & \multicolumn{3}{c}{V$_{atom}$}\\
\multicolumn{1}{c|}{(grid points/\AA)}& $\Delta^{10}_{20}$ & $\Delta^{20}_{30}$  & \\[1.5mm] \hline\\[-1mm]
$10$ & -0.00617 & -0.00112 & \\[1.5mm]
$20$ & -0.00612 & -0.00111 & \\[1.5mm]
$30$ & -0.00610 & -0.00111 & \\[1.5mm]
$40$ & -0.00611 & -0.00110 & \\
\end{tabular}
\end{ruledtabular}
\end{table}
\subsection{Charge density grids V$_{sys}$ and V$_{atom}$}\label{ssc:chargedensitygrid}
\indent As was mentioned in the previous section, the use of stored grid-based CDDs introduces small inaccuracies due to the need for interpolation between the existing grid points. The charge of the C atom in a CO molecule is shown in Fig.~\ref{fig:HIR_GridConv} as a function of the grid spacing used in the V$_{sys}$ grid. The different curves are for different grid spacings used in the V$_{atom}$ grids, from which the atomic radial CDD $\rho_{C}^{i}(n_{C}^{i},\mathbf{r})$ and $\rho_{O}^{i}(n_{O}^{i},\mathbf{r})$ are generated. It clearly shows the influence of both grids to be independent, since all curves have the same shape. Looking in detail at the exact numbers reveals that for both grids the same accuracy is obtained (\textit{cf.}~Table~\ref{table:gridConv}). This means that the change in population of the C atom in CO is the same when the same changes are made in either the V$_{atom}$ or the V$_{sys}$ grid; \textit{i.e.} the change in the population (in absolute value) of the C atom is comparable when going from the black to the red curve and when going from a point at $10$ grid points per \AA\ to a point at $20$ grid points per \AA\ on the same curve in Fig.~\ref{fig:HIR_GridConv}. Figure \ref{fig:HIR_GridConv} and Table~\ref{table:gridConv} also show that quite a dense mesh is needed to obtain very accurate results. Though this is not a big problem for periodic systems with small unit cells, it could become problematic for molecules which require big unit cells to accommodate the vacuum required to prevent interaction between the periodic copies. The same is true for high accuracy V$_{atom}$ grids, which are required for high accuracy atomic radial densities. Fortunately, these must only be generated once, and the high resolution radial densities can then be stored in a library.\\
\begin{figure}[tb!]
  \includegraphics[width=8cm,keepaspectratio=true,trim=5 0 5 5,clip]{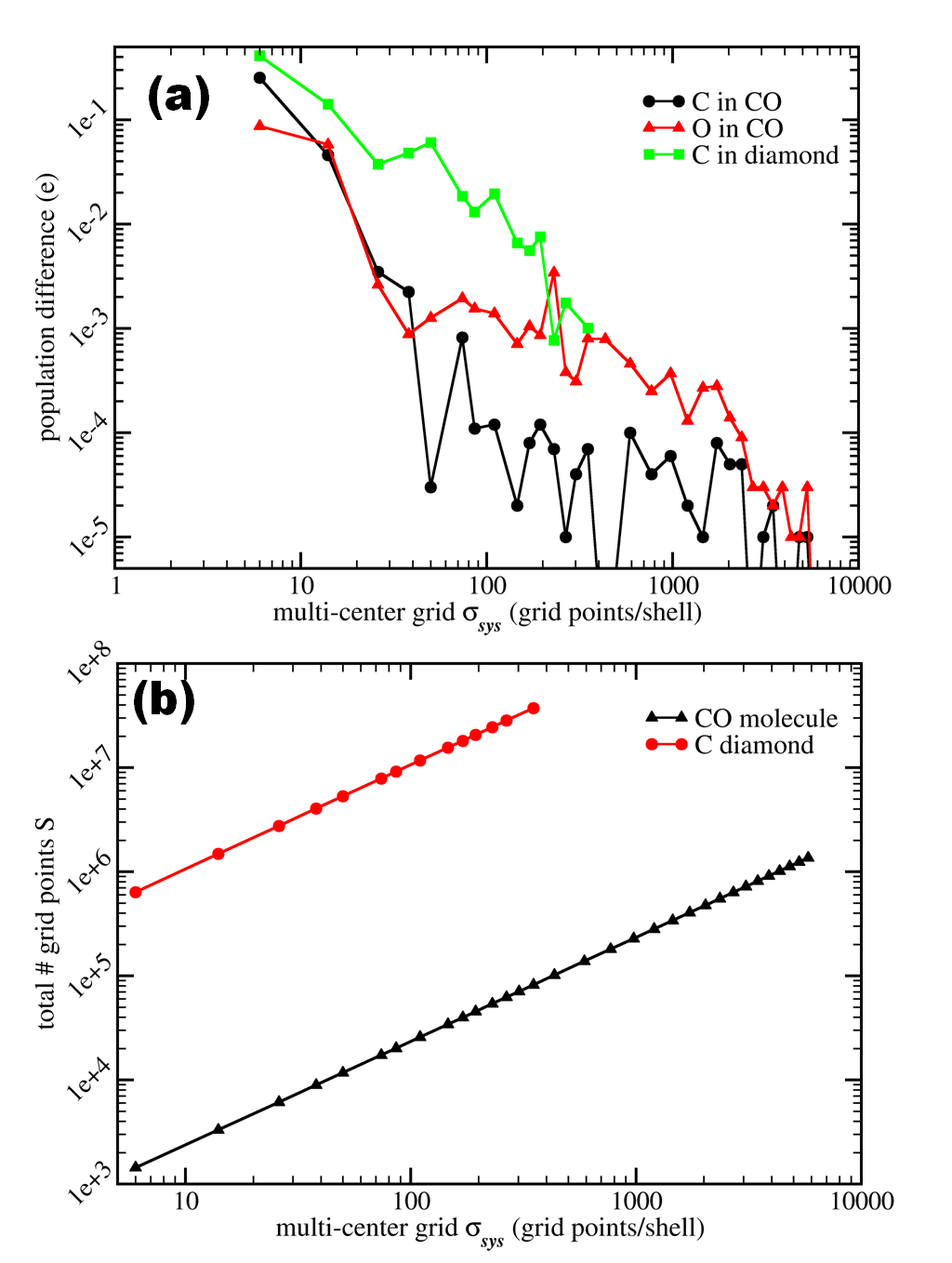}\\
  \caption{(a) Convergence behavior of the population/charge as function of the number of grid points per spherical integration shell. For the C(black line) and O(red line) atom of the CO molecule the value shown is the absolute value of the difference between the calculated population and the calculated population using the most dense grid. In case of the C atom in the diamond system (green line) the charge should be zero, so the absolute value of the calculated charge is presented. (b) The total number of grid points as function of the number of Lebedev-Laikov grid points per spherical shell.\cite{LebedevVI_grid:1999DokladyMath} The large number of grid points for the diamond system, is due to its large SoI (see text).}\label{fig:HirI_nlebConv}
\end{figure}
\subsection{Spherical integration grids S$_{sys}$ and S$_{atom}$}\label{ssc:sphereIntgrid}
\indent Because the populations and charges in the Hirshfeld(-I) approach are obtained by an integration of the CDD of the system, attention needs to be paid to the integration grid used. The current implementation uses a multi-centered grid, as was noted in Sec.~\ref{ssc:spatInteg}. In this setup we used Lebedev-Laikov grids on the spherical shells of the integration grids.\cite{LebedevVI_grid:1999DokladyMath} Figure~\ref{fig:HirI_nlebConv}a shows the influence of the number of grid points per spherical shell on the accuracy; note the log-log scale used. In general, a denser grid results in a more accurate value for the population. However, different atoms in the same system, and the same calculation can show different convergence, as is shown by the curves of the C and O atoms of a CO molecule. To have the population of the system atoms converged to within $0.001$ electron a few hundred grid points per shell are required. For a molecular calculation the multi-center grid, S$_{sys}$, only requires a few tens of thousands of grid points, as is seen in Fig.~\ref{fig:HirI_nlebConv}b. However, a Hirshfeld-I calculation for a bulk material such as diamond, which also has only two atoms in its unit cell, requires a multi-center grid S$_{sys}$ with several tens of millions of grid points. This difference by a factor thousand originates from the fact that the SoI for the diamond unit cell contains a few thousand atoms, which all contribute to the total number of grid points that need to be evaluated. This makes it very important to reduce the SoI to an as small as possible size without significant loss of accuracy.

\begin{figure*}[t!]
  \includegraphics[width=18cm,keepaspectratio=true]{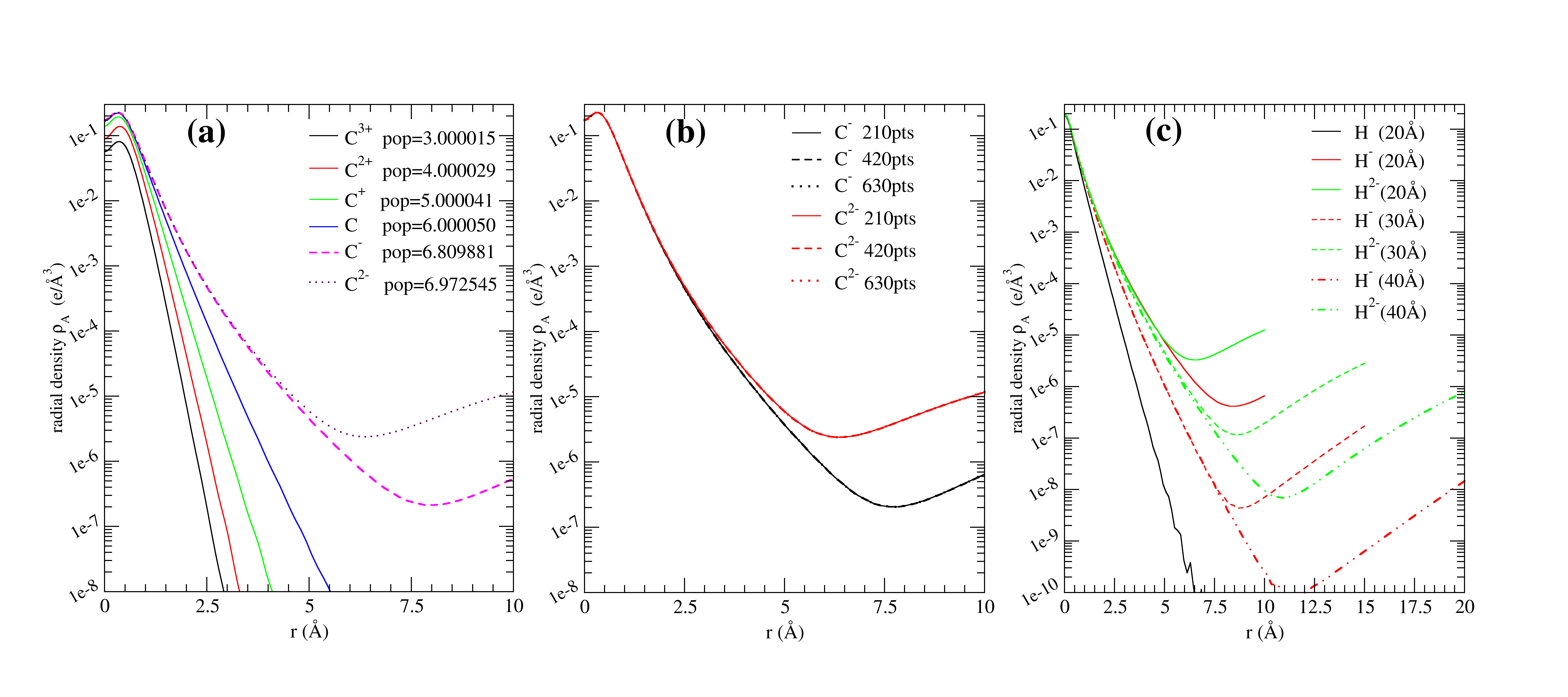}\\
  \caption{Atomic radial charge densities $\rho_{A}(r)$ calculated from grid stored atomic charge distributions. (a) Radial  charge densities for different carbon ions obtained from charge distributions in a cubic periodic unit cell of $20.5\times 20.5\times 20.5$\AA$^3$. The populations (pop) given of each ion are calculated by spherical integration of this radial distribution. (b) Radial charge distributions for two negative carbon ions obtained from the same size periodic unit cell as (a), but with different grid spacings: $210$, $420$, and $630$ grid points per $20.5$\AA. (c) Radial CDDs for different hydrogen ions. For the H$^-$ and H$^{2-}$ ions the results for different unit cell sizes are compared: $20\times 20\times 20$\AA$^3$, $30\times 30\times 30$\AA$^3$, and $40\times 40\times 40$\AA$^3$.}\label{fig:radialdens}
\end{figure*}
\subsection{Atomic radial CDDs}
\indent The atomic CDDs are calculated in a periodic cell under PBC, and stored on $3$D charge density grids V$_{atom}$ (\emph{cf.}~Sec.~\ref{ssc:chargedensitygrid}). The atomic radial CDDs are obtained from these through spherical averaging of the density distributions. Spherical averaging is done using the S$_{atom}$ grid,
with spherical shells containing Lebedev-Laikov grids of $5810$ grid points.\cite{LebedevVI_grid:1999DokladyMath} The resulting distributions for different C ions are shown in Fig.~\ref{fig:radialdens}a. The populations obtained by spherical integration over these distributions shows that the correct populations are obtained for the neutral and the positively charged ions. The negatively charged ions show a population which is too small. Moreover, the curves in Fig.~\ref{fig:radialdens}a increase again for longer ranges. This increase is not (only) due to overlapping tails of periodic copies in neighboring cells; extrapolating the decreasing part of the curve, and multiplying with the number of periodic nearest neighbors would give a much smaller value at a distance of $10$\AA\ than is now the case. The origin of the increase lies in the fact that the plane wave approach used for the atom calculations can only bind a limited amount of extra electrons to a given atom. This amount varies from atom to atom. As a result, it tries to place the excess electrons as far from the atom as possible. Due to strong delocalization inherent to plane waves these electrons are spread out over the vacuum between the atoms, with the highest electron density at the center of the unit cell; \textit{i.e.}~as far from the atoms as possible.\\
\indent It is interesting to note that this artifact is purely due to the use of plane waves, which try to smear out the unbound electrons over the entire (empty) space. If one uses Gaussians instead, the additional electrons are
bound by definition through the basis set used, even if these electrons should not be bound to the atom anymore. From the modeling point of view one might prefer this latter type of artifact over the former, since we are interested in the CDD of the electrons for `free ions', irrespective of their bound or unbound nature. Later in this section we show how this delocalization problem can be solved in a simple way.\\
\indent Figure~\ref{fig:radialdens}b shows that this artifact is independent of the resolution used for the V$_{atom}$ grid, as the different curves nicely overlap. Figure~\ref{fig:radialdens}c shows the influence of the periodic cell size on this artefact. In case of the presented hydrogen ions, it shows that using a cubic periodic cell with a side of $20$\AA, the curves for H$^-$ and H$^{2-}$ coincide in the short range region. Moreover, they don't show the expected exponential decay, and increase sharply at longer range. Using a larger unit cell appears to solve these problems: firstly, the radial distributions for the two ions become distinguishable, and show the expected exponential decay. Secondly, the point where the excess electrons start to interfere noticeably is pushed back to a larger distance.\\
\indent This type of behavior is seen for all atom types investigated, positive up to neutral ions give the expected radial distributions, while the negative ions seem hampered by the fact that only a fraction of the additional electrons can be attached to the atom. Fluorine and chlorine are in this respect exceptional since for these atoms also the F$^-$ and Cl$^-$ ions give good distributions and populations. This could be considered a result of the high electron affinity of these elements.\\
\indent We find that for most negatively charged ions the populations are too small. As a result, the calculated Hirshfeld weight \wH\ for a negatively charged AIM A is underestimated. The easiest way to compensate for this discrepancy is by scaling these specific distributions such that the correct population is found after integration. This way the shape of the curve is maintained, but the resulting weights \wH\ increase. To investigate the effects of such a scaling and the erroneous tails shown in Fig.~\ref{fig:radialdens},
we compare the results of three types of atomic radial density distributions.
To that end, the reference set of molecules previously used in Hirshfeld-I studies by Bultinck \textit{et al.}\cite{BultinckP:2007JCP_HirRef,VanDamme_Bultinck:JCTC2009} is used. This set consists of $168$ neutral molecules containing only H, C, N, O, F, and Cl.\\
\indent The first set of atomic radial CDDs, called R1, contains the density distributions as shown in Fig.~\ref{fig:radialdens}a, where the radial distribution is obtained from a periodic cell of $20\times 20\times 20$\AA$^3$.\\
\indent For the second set, which is referred to as R2, we have combined these results with the results from a periodic cell of $40\times 40\times 40$\AA$^3$ but with a lower grid resolution. In this case, the core part of the radial distribution is taken from the $20\times 20\times 20$\AA$^3$ unit cell with the high resolution grid and connected to the tail part obtained from the $40\times 40\times 40$\AA$^3$ unit cell. As a result, the high accuracy for the core part of the distribution is maintained, and the tail is corrected through the removal of the delocalized electron contribution. Note that for these distributions, the curves are limited to a distance of $10$\AA\ from the core, i.e. the same maximum radius as is available from the $20\times 20\times 20$\AA$^3$ periodic cells. Because the excess tail electrons are not included anymore, the spherically integrated populations of the negative ions are slightly smaller than they are for the R1 set.\\
\indent For the third set, R3, the same procedure as for R2 is used, but this time the curves are normalized for the negative ions such that the correct population is given on spherical integration of these radial CDDs.\\
\indent To test the accuracy of the results obtained in our periodic implementation using these three different atomic radial density distribution sets, we compare the results for a large benchmark set of molecules with those obtained by a Hirshfeld-I implementation based on a molecular program (\textit{cf.}~Sec.~\ref{ssc:Hir_Methods_solstatecodes}).\\
\begin{figure}[tb!]
  \includegraphics[width=8cm,keepaspectratio=true,trim=5 0 3 3,clip]{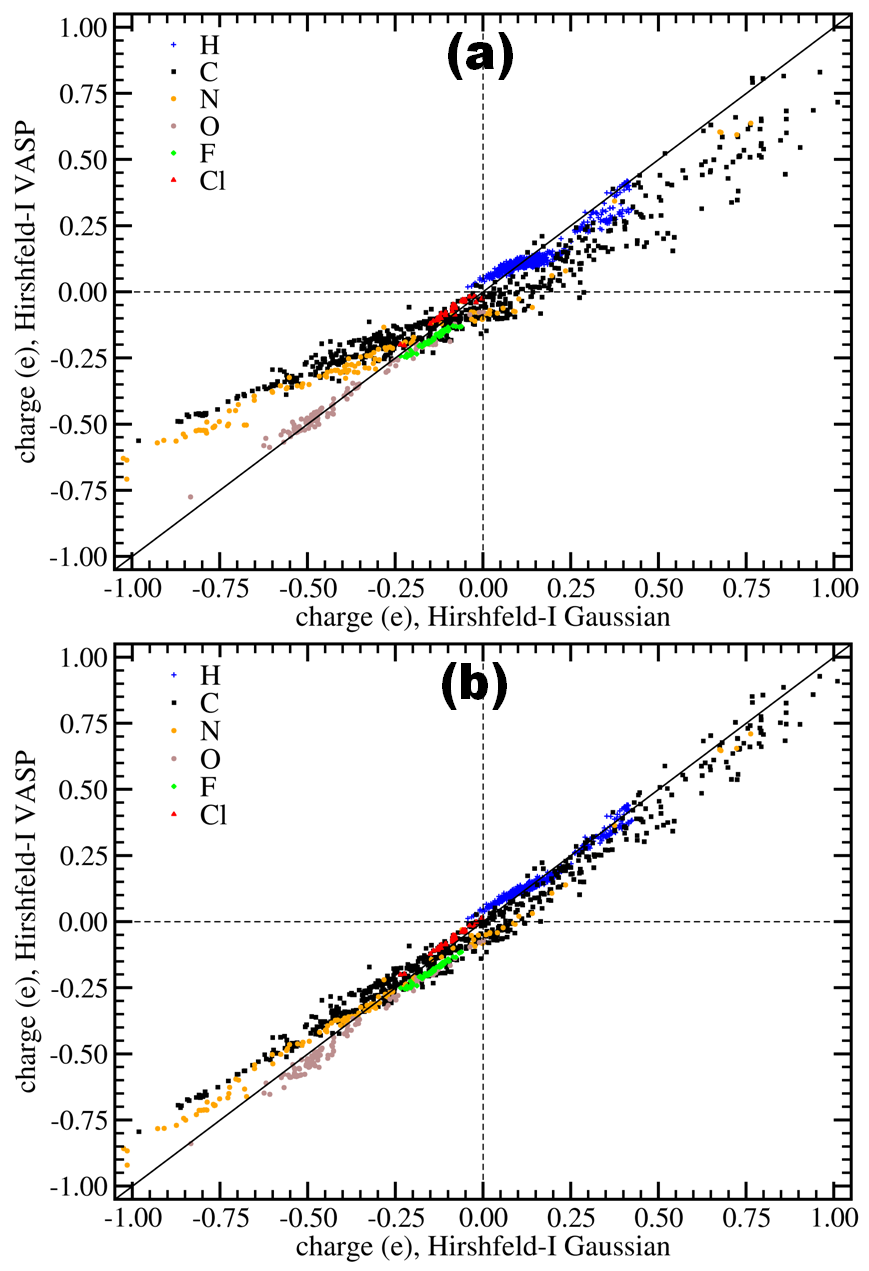}\\
  \caption{Correlation plots of the charges obtained using the periodic Hirshfeld-I implementation and the standard molecular implementation. (a) using the atomic radial density distributions of set R2 with the corrected tails, and (b) those of set R3 with the corrected tails and correctly normalized distributions (see text). In both cases the VASP optimized geometry was used for the VASP based Hirshfeld-I calculations, while Gaussian optimized geometries were used for the reference calculations.}\label{fig:HirI_CompGvsV}
\end{figure}
\indent First, two sets of calculations are performed to check the influence of the geometry on the results. The first set used the optimized geometries obtained from Gaussian calculations, and the second set used optimized geometries obtained through \textsc{VASP} calculations. For both sets the R1 atomic radial distributions are used. Table~\ref{table:VASPvsGaussianCorr} shows that the resulting correlations are nearly identical, despite the small differences in geometries for the two sets. Although the correlation coefficients $R^2$ can be considered reasonable, the spread on the C and N data points is quite large. Furthermore, the slope of the linear fit of the C and N data is much too big. In both cases the intercepts are acceptably small.\\ \indent Looking at the effect of fixing the tail of the atomic distributions, through the use of the R2 set of atomic radial density distributions, we see a slight increase in the slope for the C and N data, but in general the obtained values for the correlation, intercept and slope remain comparable to the previous sets of calculations. For the results obtained with the R2 radial distributions, Table~\ref{table:VASPvsGaussianCorr} also shows the standard deviation of the difference between our calculated results and those of the reference data. The values are similar for the first two sets, which are therefore not shown. Table~\ref{table:VASPvsGaussianCorr} shows the deviation for the C and N data sets to be one order of magnitude larger than for the other elements. It is unclear to the authors why specifically these two elements show such a bad behavior. Looking at the underestimation of the atomic population for the monovalent anions we find an error of roughly $0.5$e for H and N, and $0.3$e for C and O. On the other hand for F and Cl an underestimation of $0.1$ and $0.0$e, respectively, is found. The only aspect in which C and N differ from the other elements is that both can show large positive \emph{and} negative charges.\\
\indent The correlation results of the R3 atomic radial density distributions show a clear improvement over the previous results. For each data set the slope is closer to unity, though the intercepts remain as before. Especially the C and N results show a large improvement. Their standard deviation gets halved which is clearly visible in the correlation plots in Fig.~\ref{fig:HirI_CompGvsV}a and b. This immediately shows that the simple scaling used for fixing the underestimated population due to the delocalization problem does not introduce large artifacts. In contrast, it actually shows that the shape of the radial CDD of the negatively charged ions is correct.\\

\begin{table*}[tb!]
\caption{The fitting and correlation results for the different sets of radial CDDs used in Hirshfeld-I calculations for a set of $168$ molecules. The molecular geometries are either optimized with Gaussian or \textsc{VASP}. The radial distributions R1 are the default distributions obtained from VASP atomic density distributions. The radial distributions R2 contain a fix for the tail of the distribution, and the radial distributions R3 contain a fix for the tail of the distributions and in addition the resulting distributions are normalized to give the correct number of electrons (see text). $a$ and $b$ are the slope and the intercept of the linear fit. $R^2$ gives the correlation with Gaussian molecular results, and $\sigma$ gives the standard deviation. \label{table:VASPvsGaussianCorr}}
\begin{ruledtabular}
\begin{tabular}{l|crc|crc|crcc|crcc}
& \multicolumn{3}{c|}{Gaussian geometry} & \multicolumn{11}{c}{\textsc{VASP geometry}}\\
& \multicolumn{3}{c|}{R1} & \multicolumn{3}{c|}{R1} & \multicolumn{4}{c|}{R2}& \multicolumn{4}{c}{R3}\\
\hspace{0.75cm} & $a$ & \multicolumn{1}{c}{$b$} & $R^2$ & $a$ & \multicolumn{1}{c}{$b$} & $R^2$ & $a$ & \multicolumn{1}{c}{$b$} & $R^2$ & $\sigma$ & $a$ & \multicolumn{1}{c}{$b$} & $R^2$ & $\sigma$ \\
\hline
H & $1.187$ & $-0.014$ & $0.940$ & $1.208$ & $-0.019$ & $0.944$ & $1.185$ & $-0.008$ & $0.933$ & $0.026$ & $1.109$ & $-0.021$ & $0.979$ & $0.016$ \\
C & $1.353$ & $ 0.020$ & $0.966$ & $1.344$ & $ 0.025$ & $0.970$ & $1.439$ & $ 0.020$ &   $0.963$ & $0.181$ & $1.131$ & $ 0.013$ & $0.989$ & $0.086$ \\
N & $1.454$ & $ 0.026$ & $0.978$ & $1.455$ & $ 0.018$ & $0.979$ & $1.483$ & $ 0.025$ &   $0.975$ & $0.154$ & $1.161$ & $ 0.032$ & $0.995$ & $0.072$ \\
O & $1.131$ & $ 0.066$ & $0.992$ & $1.123$ & $ 0.057$ & $0.991$ & $1.152$ & $ 0.064$ &   $0.992$ & $0.031$ & $1.025$ & $ 0.048$ & $0.993$ & $0.019$ \\
F & $1.110$ & $ 0.064$ & $0.982$ & $1.150$ & $ 0.071$ & $0.977$ & $1.197$ & $ 0.070$ &   $0.980$ & $0.011$ & $1.162$ & $ 0.071$ & $0.987$ & $0.009$ \\
Cl& $1.117$ & $-0.009$ & $0.979$ & $1.124$ & $-0.018$ & $0.980$ & $1.110$ & $-0.013$ & $0.966$ & $0.015$ & $1.072$ & $-0.018$ & $0.993$ & $0.007$ \\
\end{tabular}
\end{ruledtabular}
\end{table*}

\begin{figure}[tb!]
  \includegraphics[width=8cm,keepaspectratio=true]{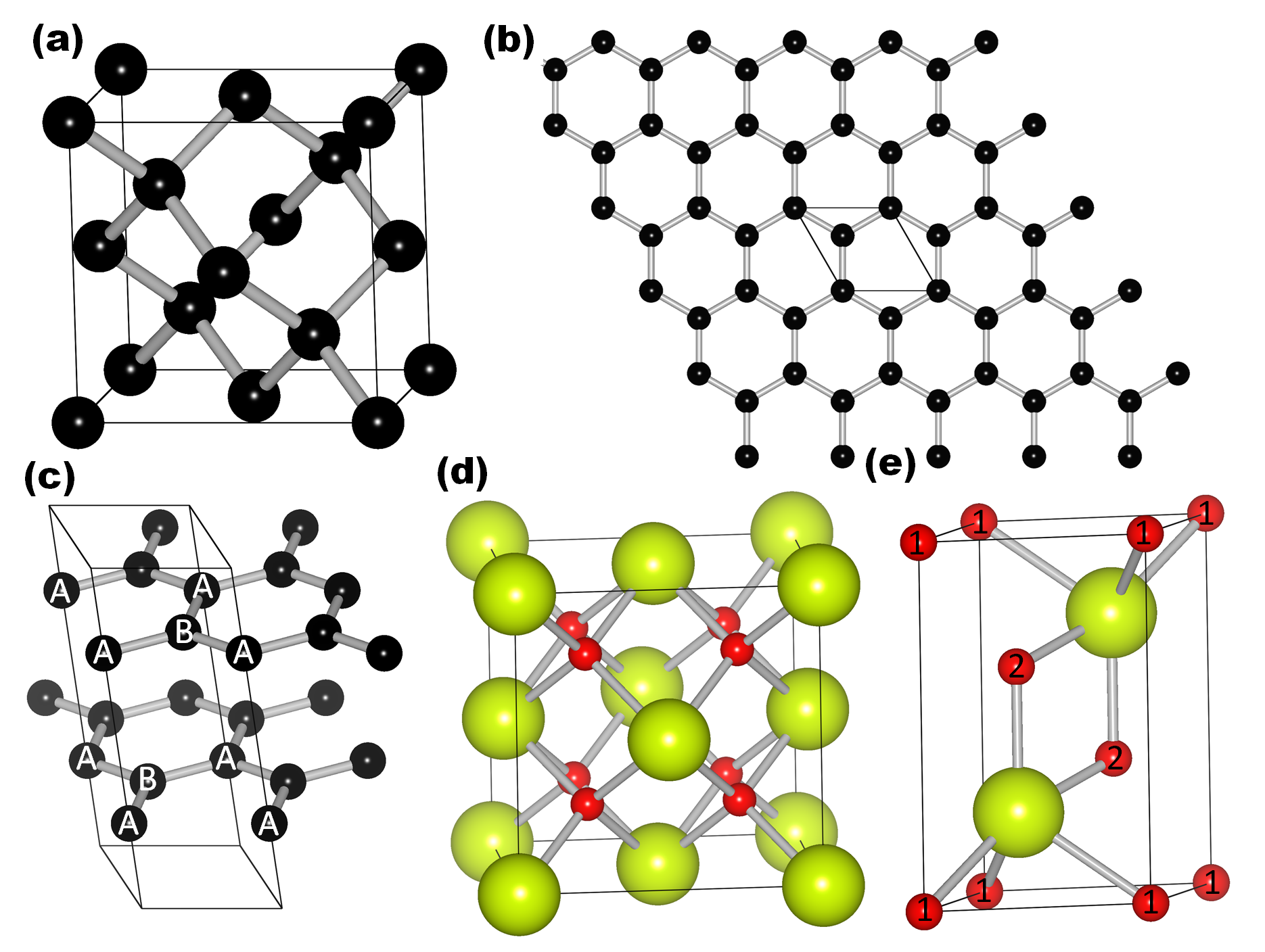}\\
  \caption{Ball-and-stick representations of (a) the cubic diamond super cell, (b) the graphene sheet (unit cell indicated with black parallelogram), (c) graphite (unit cell indicated) in black, (d) the cubic CeO$_2$ super cell, and (e) the Ce$_2$O$_3$ unit cell. The black, red and yellow spheres indicate the positions of the carbon, oxygen and cerium atoms,respectively. The inequivalent C atoms in the graphite structure are indicated as $A$ and $B$. The $B$ atoms are always located at the center of the hexagons of the neighboring sheets. The inequivalent O positions in Ce$_2$O$_3$ are indicated as $1$ and $2$.\\
  The ball-and-stick representations are generated using the \textsc{VESTA} visualization tool.\cite{VESTA:JApplCryst2008}}\label{fig:solidsGeom}
\end{figure}
\subsection{Periodic systems}
\indent In this final section some actual periodic systems are considered. Their structures are presented in Fig.~\ref{fig:solidsGeom}: the systems of choice are diamond, graphene, graphite, CeO$_2$, and Ce$_2$O$_3$.
The diamond and graphene systems, have  the property that all C atoms are equivalent, which should result in zero charges on all atoms. Graphite is quite similar to graphene, however, two inequivalent C positions are present (\textit{cf.}~Fig.~\ref{fig:solidsGeom}c). The ceria systems on the other hand are chosen for the presence of Ce atoms with different valency; tetravalent Ce in CeO$_2$ and trivalent Ce in Ce$_2$O$_3$. Note that for both systems the same Ce pseudo-potential is used. Furthermore, in CeO$_2$, the O ions are all equivalent, while in Ce$_2$O$_3$ the Ce ions are equivalent but the O ions are not, only two of them are equivalent. Furthermore, for Ce$_2$O$_3$ we consider both the ferromagnetic (FM) and anti-ferromagnetic (AF) configurations, allowing us to check how strongly different spin-configurations influence the results in this system.\\
\indent For all these systems we use the radial atomic CDDs of the R3 set, with a spherical integration grids S$_{atom}$ containing $5810$ grid points per shell. Table~\ref{table:SolidGrids} shows the $k$-point sets used for the periodic systems, the number of atoms per unit cell and the number of atoms included in the SoI. The grid point separation for the V$_{sys}$ grid for each of these systems was set to $\leq0.01$\AA. The Hirshfeld(-I) populations are calculated using spherical integration grids S$_{sys}$ with $1202$ Lebedev-Laikov grid points per shell.\\
\begin{table}[tb!]
\caption{$k$-point sets and the number of atoms in the unit cell and SoI for the periodic systems under investigation. In addition, also the total number of grid points used for the spherical integration grid are given. \label{table:SolidGrids}}
\begin{ruledtabular}
\begin{tabular}{l|r@{$\times$}r@{$\times$}rccc}
& \multicolumn{3}{c}{$k$-point set} & atoms per & atoms in & grid points \\
& \multicolumn{3}{c}{}& unit cell & the SoI & ($\times 10^6$)\\
\hline
diamond     & $21$ & $21$ & $21$ & $2$ & $6374$ & $128$ \\
graphene    & $21$ & $21$ & $1$ & $2$ & $1276$ & $22$  \\
graphite    & $21$ & $21$ & $11$ & $4$ & $4618$ & $104$ \\
CeO$_2$     & $8$ & $8$ & $8$ & $3$ & $3063$ & $69$\\
Ce$_2$O$_3$ FM & $10$ & $10$ & $5$ & $5$ & $3009$ & $77$ \\
Ce$_2$O$_3$ AF & $10$ & $10$ & $5$ & $5$ & $3025$ & $78$ \\
\end{tabular}
\end{ruledtabular}
\end{table}
\begin{table}[tb!]
\caption{Hirshfeld (H$^0$) and Hirshfeld-I (H$^i$) charges calculated using LDA generated CDDs. The geometries of the periodic systems are shown in Fig.~\ref{fig:solidsGeom} where the labels for the inequivalent atoms are given.\label{table:HirshfeldResults}}
\begin{ruledtabular}
\begin{tabular}{ll|rrc}
& \hspace{1cm} & \multicolumn{1}{c}{H$^0$} & \multicolumn{1}{c}{H$^i$} &\hspace{0.5cm}\\
& & \multicolumn{1}{c}{(e)}   & \multicolumn{1}{c}{(e)} &\\
\hline
diamond     & C & $-$0.00007 & $-$0.00007 &\\[2mm]
graphene    & C &    0.00000 &    0.00000 &\\[2mm]
\multirow{2}{*}{graphite} & C$_A$ &    0.00113 &    0.00705 &\\
                          & C$_B$ & $-$0.00115 & $-$0.00707 &\\[2mm]
\multirow{2}{*}{CeO$_2$}  & Ce    &    0.59463 &    2.79393 &\\
                          & O     & $-$0.30091 & $-$1.40056 &\\[2mm]
\multirow{3}{*}{Ce$_2$O$_3$ FM}& Ce &  0.49081 &    2.32119 &\\
                          & O$_1$ & $-$0.31318 & $-$1.61119 &\\
                          & O$_2$ & $-$0.33576 & $-$1.51663 &\\[2mm]
\multirow{3}{*}{Ce$_2$O$_3$ AF}& Ce &  0.48575 &    2.32755 &\\
                          & O$_1$ & $-$0.30825 & $-$1.63062 &\\
                          & O$_2$ & $-$0.33317 & $-$1.51378 &\\
\end{tabular}
\end{ruledtabular}
\end{table}
\indent The resulting Hirshfeld (H$^0$) and Hirshfeld-I (H$^i$) charges for the inequivalent atoms are shown in Table~\ref{table:HirshfeldResults}.
It clearly shows the H$^0$ values are closer to zero (i.e. the charge at which the atoms are initialized) than the H$^i$ ones. This is the expected behavior and its origin was discussed earlier by Ayers\cite{AyersPW:2000JCP} and  Bultinck \textit{et al.}\cite{BultinckP:2007JCP_HirRef} The diamond and graphene charges are (nearly) zero as one would expect based on symmetry arguments. This shows there are no significant artifacts which introduce spurious charges due to the PBC. The results for graphite are somewhat remarkable. Table~\ref{table:HirshfeldResults} shows there is a small charge transfer going from the A to the B sites. This could be understood as a consequence of the very weak bonding between the A sites in different sheets. For each C atom, three electrons are placed in hybridized $sp^2$ orbitals, where the fourth electron delocalizes in distributed $\pi$ bonds. For the A site C atoms, the contribution to the AIS charge of these $\pi$ bond electrons is shared between the A sites of neighboring sheets. Since the C atom at the B site has no direct neighbor on the neighboring sheet the contribution goes entirely to this atom, resulting in the slightly negatively charged C atom. Charge neutrality results in a slightly positively charged C atom at the A site. Similar behavior was observed by Baranov and Kohout\cite{KohoutMBaranovA:2011JComputChem} using the Bader approach. These authors, however, find a larger and opposite charge transfer, resulting in a charge of $+0.08$e and $-0.08$e on the C$_{A}$ and C$_{B}$ atoms respectively. This difference could originate from the different methods used.\\
\indent The ceria compounds show the behavior expected with regard to equivalent/inequivalent atoms. The Hirshfeld-I values presented in Table~\ref{table:HirshfeldResults} are comparable with Bader charges presented in literature. Castleton \emph{et al.} found for CeO$_2$ Bader charges of $+2.3$e and $-1.15$e for Ce and O, respectively.\cite{castleton:jcp2007}
The Mulliken atomic charges for Ce$_2$O$_3$ presented in literature appear strongly dependent on the functional used, varying from $+1.29$e for PBE up to $+2.157$e for Hartree-Fock.\cite{hay:jcp2006,YangJDolgM:TheorChemAcc2005}
This makes it difficult to make a qualitative assessment of our obtained results.\\
Table~\ref{table:HirshfeldResults} also shows there is a clear difference between the tri-and tetra-valent Ce ions, also the different configurations for the O ions show distinctly different charges. Looking at the relative atomic charges of Ce and O atoms in CeO$_2$ and Ce$_2$O$_3$ we find the same relative order as was found by Hay \textit{et al.}\cite{hay:jcp2006} and a difference in atomic charge for the Ce ions of comparable size. The different charges for the tri-and tetra-valent Ce ions might tempt one to consider these charges as indicators of the oxidation state if not the actual oxidation state of the atoms involved. As a result one could then assume that the same charge in a different configuration would be the result of the same oxidation state (\textit{cf.}~concept of transferability). Looking at the charges of the O atoms in both CeO$_2$ and Ce$_2$O$_3$ shows this is clearly not the case, since all O atoms formally have the same oxidation state, while the calculated Hirshfeld-I charges vary $0.2$ electron. The Hirshfeld charges on the other hand show a much smaller variation of only $0.03$ electron. At this point, it is important to stress that atomic charges do not, as opposed to what is often assumed, directly reveal the oxidation state, nor the valence of an atom. A Hirshfeld(-I) analysis, just like a Bader analysis, can only reveal atomic domains. The actual valence of an atom can be derived from the localization indices, which correspond in the simplest form to integrating twice the exchange-correlation density over the same atomic domain.\cite{BaderRFWStephens:JAmChemSoc1975,PonecRCooperD:FaradDisc2007}
(De)localization indices are obtained from double integration of the exchange correlation density, which requires density matrices, in the simplest case only first order density matrices.\cite{BultinckPatrick:2010JPCA} Such matrices have been less thoroughly explored in solid state calculations.\cite{KohoutMBaranovA:2011JComputChem}\\
\indent Another interesting point to note is that different spin-configurations have little to no influence on the obtained charges. This is seen when comparing the FM and AF configurations of Ce$_2$O$_3$. This means that for generating the required system CDDs generally non-spin-polarized calculations suffice for the studied systems. Note, however, that the single atom calculations are spin-polarized.


\section{Conclusion}\label{sc:Hir_conclusions}
\indent We have presented an implementation of the Hirshfeld-I method specifically aimed at periodic systems, such as wires, surfaces, and bulk materials. Instead of calculating the charge densities at each point in space on the fly using the precalculated wave function of the system, we interpolate the charge density from a precalculated CDD on a dense spatial grid, speeding up the calculation of the density significantly. The use of such grids is possible because PBC allow for the use of a relatively small grid to describe the entire system accurately.\\
\indent Unlike total energy calculations, the number of atoms involved can not be fully reduced to only those in the unit cell. Although, the populations only need to be calculated for the atoms in the unit cell, the Hirshfeld-I calculations require a large `\emph{sphere of influence}' containing a few thousand of atoms. Using multi-center integration grids, the computational cost can be seriously reduced by cleverly selecting only the grid points involved in the calculations.\\
\indent We have shown that the uniform grids used to store both the atom and the system CDDs have an equal influence on the accuracy of the final Hirshfeld-I calculated populations, leading to the suggestion of building the library of atomic radial CDDs using as dense as possible grids. In addition, we have shown that both different atomic types and different chemical environments give rise to a different convergence behavior as function of the spherical integration grid.\\
\indent The problems observed for the atomic radial CDDs of negatively charged ions are solved in a simple way, and we show that the introduced scaling of the distributions significantly improves the obtained results for the Hirshfeld-I charges. The resulting values for a benchmark set of $168$ neutral molecules show very good agreement with the values obtained by a previous implementation of the Hirshfeld-I method aimed solely at molecular systems.\\
\indent In the final section we have investigated some periodic systems to show the validity of our implementation. For each of these systems the expected behavior of the charges is observed. Because of their simplicity these systems are ideal test cases for Hirshfeld-I implementations for periodic systems.\\
\section{Acknowledgement}
\indent The research was financially supported by FWO-Vlaanderen, project n$^{\circ}$ $3$G$080209$. This work was carried out using the Stevin Supercomputer Infrastructure at Ghent University, funded by Ghent University, the Hercules Foundation and the Flemish Government--department EWI.


\bibliography{danny,Hirshfeld,notes}

\begin{thebibliography}{52}
\bibitem[{fn:()}]{fn:CHGCAR}
\bibinfo{note}{In our setup we use the {CHGCAR}-file produced by VASP as input.
  Due to it's simple format this means that comparably formatted input can
  easily be generated for different solid state codes, making our setup
  independent of the used solid state/quantum chemical code}.
\end{thebibliography}

\end{document}